\documentclass[12pt]{article}
\usepackage{amsmath}
\usepackage{amsthm}
\usepackage{graphicx,psfrag,epsf}
\usepackage{enumerate}
\usepackage{natbib}
\usepackage{booktabs}
\usepackage{url}

\newcommand{\blind}{0}

\begin{document}

\newtheorem{theorem}{Theorem}[section]

\bibliographystyle{abbrvnat}

\def\spacingset#1{\renewcommand{\baselinestretch}%
{#1}\small\normalsize} \spacingset{1}


\if0\blind
{
  \title{\bf Weighted composite likelihood for linear mixed models in complex samples}
  \author{Thomas Lumley\thanks{Part of this research was supported by the Marsden Fund Council from New Zealand Government funding, managed by Royal Society Te Ap\=arangi.}\hspace{.2cm}\\
    Department of Statistics, University of Auckland\\
    and \\
    Xudong Huang \\
     Department of Statistics, University of Auckland\\}
  \maketitle
} \fi

\if1\blind
{
  \bigskip
  \bigskip
  \bigskip
  \begin{center}
    {\LARGE\bf Weighted composite likelihood for linear mixed models in complex samples}
\end{center}
  \medskip
} \fi

\bigskip
\begin{abstract}
Fitting mixed models to complex survey data is a challenging problem.  Most methods in the literature, including the most widely used one, require a close relationship between the model structure and the survey design. In this paper we present methods for fitting arbitrary mixed models to data from arbitrary survey designs. We support this with an implementation that allows for multilevel linear models and multistage designs without any assumptions about nesting of model and design, and that also allows for correlation structures such as those resulting from genetic relatedness. The estimation and inference approach uses weighted pairwise (composite) likelihood.
\end{abstract}

\noindent%

\spacingset{1.45}
\section{Introduction}
\label{sec:intro}

Fitting mixed models --- even linear mixed models --- to data from multistage surveys is a challenging problem.  An initial difficulty is  ``there is nowhere to stick the weights''; the Gaussian multivariate loglikelihood is not a sum of single-observation terms. Another aspect of the problem is that mixed models are fundamentally about bias:variance tradeoffs. Standard design-based inference is about `grossing-up', reweighting the data to look like the population (or other sampling frame), but the bias:variance tradeoff we care about is the one in the sample, not the one in the population. 

Our research was initially motivated by the problem of fitting quantitative trait locus models to data from the Hispanic Community Health Study/Study of Latinos.  HCHS/SoL is a cohort study of Hispanic and Latino Americans, with recruitment based on a multistage probability sampling scheme that sampled census block groups, then households, then individuals within households, in six sites chosen for representation of varying Hispanic and Latino populations \citep{hchs-design}. When fitting a linear mixed model to these data, the relatedness structure is not nested in the sampling structure: people in the same household can be unrelated (eg, spouses) and people in different households or block groups can be related. 

We are interested in methods that estimate both the variance components and the fixed-effects regression parameters.  The primary existing methods were those of \citet{pfeffermann-mixed} and \citet{rabe-hesketh-survey}. They assumed that the structure of the multilevel model and the multistage sample were matched, so that each group in the model could be assigned to a sampling unit in the survey.  Reweighting is then possible using the conditional independence of sampling at each stage of the design and of random effects at each level of the model.  This stagewise reweighting approach does not immediately give good results for small clusters, because the weighted and unweighted problems have different bias:variance tradeoff.  Simple rescaling of the weights gives excellent results under reasonable sampling designs, though unreasonable sampling designs can cause significant bias \citep{svylme-stat}.  Stagewise reweighting can also be extended to generalised linear mixed models using adaptive Gaussian quadrature.  There is a widely used implementation in Stata \citep{stata}, first as the {\tt gllamm} package \citep{gllamm-framework} and now as part of the standard program. This approach is also implemented in Mplus \citep{mplus} and MLwiN \citep{mlwin}.

Rao and co-workers \citep{rao-mixed,yi-rao-mixed} proposed an approach using weighted pairwise likelihood. Their approach, like the stagewise reweighting approach, required a close relationship between the sampling design and the model structure, and they did not publish an implementation. An advantage of weighted pairwise likelihood is that it gives consistent and asymptotically Normal estimators under asymptotics where either the number or the size of clusters increases.

 In the discussion of \citet{pfeffermann-mixed}, Rao and Roberts had already pointed out that the required relationship between the design and the model was restrictive:
 \begin{quote}
 The authors assume that the sample is selected according to the hierarchical structure of the model,  but in multipurpose surveys the hierarchical structure of the sample could be quite different.
 \end{quote}
 
 In this paper we relax those requirements and allow essentially any linear mixed model and design. We also show that using all pairs in the composite likelihood, rather than just correlated pairs, may sometimes be advantageous and is not computationally prohibitive.  We present an implementation for R \citep{r-itself} in the package {\tt svylme} \citep{svylme-pkg}.  

In section~\ref{sec:meth} we discuss the inferential approach and some computational issues for implementation.  In section~\ref{sec:eg} we give an examples based on a survey data sets. We also analyse two examples of complete unweighted data to separate issues with pairwise likelihood estimation from issues with weighting. Section~\ref{sec:verify} presents results from simulation experiments showing properties of the estimators, including samples using the data sets in section~\ref{sec:eg} as populations.  These simulations examine the performance of the pairwise likelihood estimator both when the variance components are of substantive interest and when  the regression coefficients are of primary interest. All the code and data are in the Supplemental Materials. We have previously published simulations for the setting where the design and sampling units are the same and random effects are independent \citep{svylme-stat}, so we focus here on correlated random effects or designs not following the model structure. Finally, in section~\ref{sec:conc} we discuss findings about efficiency of the estimators and challenges for implementation in generalized linear mixed models.

\section{Methods}
\label{sec:meth}

We use the Laird--Ware formulation of the linear mixed model \citep{laird-ware-mixed}. For an individual $i$ we observe an outcome  vector $Y$ and predictor matrices $X$ and $Z$ satisfying
\begin{equation}
Y=X\beta+Zb+\epsilon
\end{equation}
where $X$ are the fixed-effect predictors, $Z$ are the random-effect predictors, 
$\epsilon\sim N(0,\sigma^2)$ and $b\sim N(0,\sigma^2 V(\nu))$. We write $\theta=(\beta,\sigma, 
\nu)$ for the full parameter vector, and $p$ and $q$ for the dimensions of $X$ and $Z$ respectively. Under this model, $Y$ is multivariate Gaussian with mean vector $\mu=X\beta$ and covariance matrix 
\begin{equation}
\sigma^2\Xi = \sigma^2(I+Z^TV(\nu)Z).
\end{equation}

In our implementation we allow arbitrary correlation between different random effects on the same unit -- eg, between intercept and slope random effects. We allow structured correlation between random effects on different units; for example, correlation proportional to genetic relatedness in a family study or agricultural experiment \citep{lme4qtl}. For example, we might model 
$$\Xi=I + \rho_e E +\rho_g \Phi$$
where $E$ is a block diagonal matrix indicating which individuals share households and $\Phi$ is a genetic relatedness matrix.  The environmental term can be modelled by a single iid random effect for each household, and the genetic term by taking iid random effects for each individual and multiplying them by a square root of $\Phi$.  In general, the matrix $\Xi$ must be a linear combination of basis matrices, and if these basis matrices are not made up of blocks of indicators for a grouping factor they must be specfied explicitly.  The class of random-effect structures differs from that in \cite{lme4qtl} in not allowing equality constraints between parameters.  The resulting covariance matrix $\sigma^2\Xi$ is typically sparse, but not block-diagonal.
More general correlation involving non-linear parameters, such as autoregression over time or spatial processes, are supported by the methods but not by the current implementation.

Our goal is design-based inference about the mixed model.  That is, we take the population of size $N$ as having been generated by sampling from the mixed model, but we observe data only on a well-defined probability sample of size $n$ from the population.  We want to draw inference about the population and the model that generated it, so our target of inference does not depend on the sampling design.

We write $R_i$ for the indicator that individual $i$ in the population is sampled, and $\pi_i=E[R_i]$ for the sampling probability.  We will also need pairwise sampling indicators $R_{ij}=R_iR_j$ and pairwise sampling probabilities $\pi_{ij}=E[R_{ij}]$.  The sampling probabilities must be non-zero for all individual observations and pairs of observations in the population and must be known for all observations and pairs of observations in the sample.  

The sample will often have been taken by stratified sampling.  In this process, the population is partitioned into $K$ groups, called strata, with stratum $k$ containing $N_k$ smaller groups, called primary sampling units. A prespecified number $n_k$ of sampling units is sampled with equal probabilities and without replacement from  stratum $k$.  In multistage stratified sampling this procedure is then repeated recursively to sample within each primary sampling unit. It is straightforward to compute the resulting probabilities $\pi_i$ and $\pi_{ij}$ from the $n_k$ and $N_k$ at each stage. 

\subsection{Composite likelihood}
Composite likelihood, proposed by \citet{lindsay-composite} and reviewed by \citet{composite-review}, is an objective function constructed as a product of terms that are themselves genuine likelihoods but are not necessarily independent.  In our case, we are interested in the pairwise likelihood or its logarithm.

For any pair of observations $(i,j)$ we can easily compute the Gaussian loglikelihood $\ell_{ij}(\theta)$.  Since this is a genuine loglikelihood, its derivative has zero mean at the true parameter values. Adding up $\ell_{ij}$ over either all pairs $(i,j)$ or all pairs where $Y_i$ and $Y_j$ are not independent gives an objective function $\ell^{(P)}(\theta)$ that we call the pairwise loglikelihood. By linearity of expectation, the derivative of the pairwise loglikelihood also has zero mean at the true parameter values; the pairwise score equations are unbiased estimating equations.  A pairwise likelihood has been used previously for generalised linear mixed models and related models outside the survey context by various authors \citep{heagerty-lele,varin-vidoni-pairwise, liu-pairwise-thesis,vasdekis-ordinal,vasdekis-pairwise}.

We use an $ij$ subscript for scalars indexed by pairs of observations, eg, $\pi_{ij}$ and $[ij]$ for vectors of length 2 or $2\times 2$ matrices for a pair, eg $Y_{[ij]}$, $\Xi_{[ij]}$. We write ${\cal S}=\{(i,j): \pi_{ij}\neq \pi_i\pi_j\}$ for the set of pairs correlated under the sampling design and ${\cal P}=\{(i,j): \Xi_{ij}\neq0\}$ for the set of pairs correlated under the model. 

\subsubsection{Weighting}
In our setting, the key advantage of the pairwise loglikelihood is that it is a sum, and can thus be reweighted. Let $R_{ij}$ be  the indicator that observations $i$ and $j$ were both sampled, and let $\pi_{ij}=E[R_{ij}]$. Writing
\begin{equation}
\hat\ell^{(P)}(\theta) =\sum_{i,j\in{\cal P}}\frac{R_{ij}}{\pi_{ij}} \ell_{ij}(\theta)
\end{equation}
we have
$$E_{Y,R}\left[\frac{\partial\hat\ell^{(P)}}{\partial\theta} \right]_{\theta_0}=E_{Y,R}\left[ \sum_{i,j\in{\cal P}}\frac{R_{ij}}{\pi_{ij}} \frac{\partial}{\partial\theta}\hat\ell_{ij}(\theta)\right]_{\theta_0}=E_{Y}\left[ \sum_{i,j\in{\cal P}} \frac{\partial}{\partial\theta}\hat\ell_{ij}(\theta)\right]_{\theta_0}=0.$$
The weighted pairwise score equations are unbiased estimating equations, for the true $\theta_0$ if the model is correctly specified and for the same `least-false' $\theta^*$ as the population pairwise likelihood estimator if the model is misspecified. 

The composite likelihood does require pairwise sampling probabilities. These are straightforwardly available to a survey designer, and can be computed for multistage cluster samples from the sampling fractions at each step.  In section~\ref{sec:conc} we discuss what can be done when these are not known.

\subsubsection{Linear mixed models}
\label{sandwich}

Specialising to linear mixed models, we can profile out $\beta$ and $\sigma$ from the weighted pairwise loglikelihood in essentially the same way as for the full loglikelihood: 

$$-2\hat\ell^{(P)}(\beta,\sigma^2,\nu)= \sum_{i,j\in {\cal P}} \frac{R_{ij}}{\pi_{ij}} \left(\log|\Xi_{[ij]}(\nu)|+2\log\sigma^2+\frac{1}{\sigma^2}\left[Y-X\beta\right]_{[ij]}^T\Xi^{-1}_{[ij]}(\nu)\left[Y-X\beta\right]_{[ij]}\right) $$

Profiling out $\beta$ gives

$$-2\hat\ell^{(P)}(\sigma^2,\nu)= \sum_{i,j\in {\cal P}} \frac{R_{ij}}{\pi_{ij}} \left(\log|\Xi_{[ij]}(\nu)|+2\log\sigma^2+\frac{1}{\sigma^2}\left[Y-X\hat\beta\right]_{[ij]}^T\Xi^{-1}_{[ij]}(\nu)\left[Y-X\hat\beta\right]_{[ij]}\right) $$

Writing $\hat N=\sum_{i,j\in{\cal P}}\pi^{-1}_{ij}$ for the estimated number of pairs in the population  we now take $$\hat\sigma^2=\frac{1}{2\hat N} \sum_{i,j\in {\cal P}} \frac{R_{ij}}{\pi_{ij}}\left[Y-X\hat\beta\right]_{[ij]}^T\Xi^{-1}_{[ij]}(\nu)\left[Y-X\hat\beta\right]_{[ij]}$$
to obtain a profile weighted pairwise deviance for $\nu$ as the final objective function
\begin{equation}
\hat d_p(\nu)= \sum_{i,j\in {\cal P}} \frac{R_{ij}}{\pi_{ij}} \log|\Xi_{[ij]}(\nu)|+2\hat N\log\left( \frac{1}{\hat N}\sum_{i,j\in {\cal P}}\frac{R_{ij}}{\pi_{ij}}\left[Y-X\hat\beta\right]_{[ij]}^T\Xi^{-1}_{[ij]}(\nu)\left[Y-X\hat\beta\right]_{[ij]}\right)
\label{profile-deviance}
\end{equation}

As equation~\ref{profile-deviance} indicates, we use a generalised least squares formulation of the mixed model rather than the penalised least squared version preferred by \cite{lme4}. This choice is related to challenges in estimating the realised random effects, as discussed in section~\ref{sec:conc}.

A standard sandwich variance estimator is available for the standard errors of $\hat\beta$, since $\hat\beta$ is just a weighted least squares estimator. In particular, when the sample is much smaller than the population, we can use the usual with-replacement approximation that ignores the $O(N^{-1})$ population contribution of the variance and just computes the sampling variance. Suppose 
$$\hat\beta=(X^TWX)^{-1}(X^TWY)$$
where $W$ is an $n\times n$ matrix depending on both $\Xi$ and the pairwise sampling probabilities. Then 
\begin{equation}
\widehat{\textrm{var}}[\hat\beta]=(X^TWX)^{-1}\left(\sum_{i,j\in{\cal S}}\frac{\Delta_{ij}}{\pi_{ij}}(X^TW(Y-\hat\mu))^T_i(X^TW(Y-\hat\mu))_j\right)(X^TWX)^{-1} 
\end{equation}
where ${\cal S}$ is the set of pairs $(i,j)$ correlated by sampling, that is, the set where $\pi_{ij}\neq\pi_i\pi_j$, and $\Delta_{ij}$ is the covariance of the sampling indicators for units $i$ and $j$: $\Delta_{ij}=\pi_{ij}-\pi_i\pi_j$ if $i\neq j$ and $\Delta_{ii}=\pi_i(1-\pi_i)$.  

Calculation shows that $W_{ij}$ for $i\neq j$ depends on the off-diagonal element of the $2\times 2$ matrix $\Xi_{[ij]}^{-1}$,
$$ W_{ij}=(\Xi_{[ij]})^{-1})_{12}/\pi_{ij}$$
but $W_{ii}$ involves a sum over the $(i,i)$ diagonal element for all the pairs that $i$ contributes to:
\begin{equation}
W_{ii}= \sum_{(i,j)\in{\cal P}} \frac{(\Xi_{[ij]})^{-1}_{11}}{\pi_{ij}}.
\label{eq:sandwich-weights}
\end{equation}

In principle it is also possible to define a sandwich estimator for $\hat\nu$, involving a sum over pairs of pairs of observations, and \citet{huang} gives a lengthy proof that such an estimator is consistent. However, the estimator involves fourth-order inclusion probabilities, making it inconvenient to compute, and it appears to have poor finite-sample behaviour.  Instead, we recommend resampling approaches for inference about $\nu$.  When the sample is much smaller than the population, the usual with-replacement approximation allows a standard survey bootstrap or jackknife to be used \citep{shao-resample,raowu-boot,svrep-pkg}. The resampling approaches also gives slightly better standard error estimation for the fixed effects.

\subsubsection{All pairs or correlated pairs?}

Previous uses of the pairwise likelihood for mixed models have typically used a subset of correlated pairs \citep{rao-mixed,yi-rao-mixed, heagerty-lele, varin-vidoni-pairwise, liu-pairwise-thesis}.  Using just correlated pairs has an apparent computational advantage, and may also simplify mathematical arguments. There are settings, however, where using just correlated pairs gives poor results.  One such setting is when many observations are singletons and so appear in \emph{no} correlated pairs. For example, data on (human) births mostly comes with one baby per birth, but about 4\% of births are multiples; most individuals will be in no correlated pairs. Singleton births contribute no information about the correlation between twins, but they  contribute most of the information about the fixed effects and marginal variance.  Another such setting is data with widely varying cluster sizes, where the number of pairs in a cluster of size $m$ increases as $m^2$, potentially giving undue influence to large clusters. We give an example in section~\ref{eg:milk}. 

In our application, it is possible to compute the all-pairs composite likelihood efficiently, so the computational argument for correlated pairs disappears.  The pairwise loglikelihood for an independent pair $(i,j)$ decomposes into the sum of marginal loglikelihoods.  If all pairs were independent, the double sum of pairs would then collapse to a single sum of marginal loglikelihoods
$$\sum_{i\neq j}^N\ell_{ij}(\theta)= (N-1)\sum_{i=1}^N \ell_i(\theta)$$
For correlated pairs, we can subtract off the sum of the marginal loglikelihoods and add the correct pairwise loglikelihood. Writing ${\cal P}$ for the set of correlated pairs under the model and $\ell_i(\theta)$ for the marginal loglikelihood of observation $i$ we find in the population
\begin{equation}
\sum_{i\neq j}^N\ell_{ij}(\theta)=(N-1)\sum_{i=1}^N \ell_i(\theta)+\sum_{i,j\in{\cal P}}\left[\ell_{ij}(\theta)-\ell_i(\theta)-\ell_j(\theta)\right].
\label{eq:all-pairs}
\end{equation}
With weights we define
\begin{equation}
\hat\ell^{(P)}(\theta)=(N-1)\sum_{i=1}^N \frac{R_i}{\pi_i}\ell_i(\theta)+\sum_{i,j\in{\cal P}}\frac{R_{ij}}{\pi_{ij}}\left[\ell_{ij}(\theta)-\ell_i(\theta)-\ell_j(\theta)\right].
\label{eq:all-pairs-wt}
\end{equation}

For standard error estimation, equation \ref{eq:sandwich-weights} is modified to
\begin{equation}
W_{ii}= \sum_{j:(i,j)\in{\cal P}} \frac{(\Xi_{[ij]})^{-1}_{11}}{\pi_{ij}}+ \sum_{j:(i,j)\not\in{\cal P}}\frac{(\Xi_{ii})^{-1}}{\pi_{i}}.
\label{eq:sandwich-all}
\end{equation}

The computation time here scales as $n$ plus the number of correlated pairs, which is asymptotically smaller than for full likelihood. A cluster of size $m$ would contribute $\Theta(m^3)$ time to the full loglikelihood for matrix inverse and determinant, but contributes $O(m^2)$ pairs each taking constant time to the pairwise likelihood.

\subsection{Asymptotics}
Although the population pairwise loglikelihood is a sum, it is not a sum  over independent observations --- it is not in general even a sum over independent cluster totals --- so classical central limit theorems do not apply.   \citet{yi-rao-mixed} assumed one nested sequence of model clusters matching the sampling design. \citet{huang} relaxed the assumption to allow a sequence of nested model clusters that is unrelated to the sampling design, and to allow correlated random effects.  Here, we must further relax the assumptions  to allow multiple, potentially crossed sets of model clusters together with correlated random effects. We use an approach based on central limit theorems for sums with graph-structured dependence. The basic theorems are \citet{baldi-rinott} or \citet{janson-clt}.  A dependence graph is a graph whose vertices are observations, such that two sets of observations with no edges directly connecting one to the other are independent.   In our setting, observations will be connected by an edge if they share a random effect, or if they share a sampling unit.  For models with correlated random effects, Stein's method also provides central limit theorems under local dependence, such as those of \citet{bolthausen-stein} and \citet{guyon} under  assumptions on the strong-mixing coefficients. We give more detail in the Supplemental Materials.

\subsection{Implementation}
We use the implementation of \citet{lme4} to do data setup and provide starting values for optimisation.  For correlated random effects, we use start-up code based on \citet{lme4qtl}.  The user supplies survey data and metadata in a survey design object from the R {\sf survey} package \citep{survey-paper,survey-pkg}.  

Pairwise sampling probabilities can be supplied by the user if they are known. They can be calculated exactly by the software when the design is multistage stratified sampling, and are approximated for probability-proportional-to-size sampling using a sample-based estimate of an approximation due to H\'ajek. The population approximation is \citep[equation 9.14]{brewer}
\begin{equation}
\pi_{ij}\approx \pi_i\pi_j\left(1-(1-\pi_i)(1-\pi_j)\left(\sum_{k=1}^N \pi_k(1-\pi_k) \right)^{-1}\right)
\end{equation}
and the estimate is
\begin{equation}
\widehat{\pi}_{ij}\approx \pi_i\pi_j\left(1-(1-\pi_i)(1-\pi_j)\left(\sum_{k\in\text{sample}} (1-\pi_k) \right)^{-1}\right).
\end{equation}
 The restriction to $2\times 2$ matrices in the pairwise loglikelihood allows for some optimisations.  The determinant and inverse of a $2\times 2$ matrix can be computed efficiently by explicit formulas, and these computations can be vectorised to be efficient in interpreted R code.

We  follow \citet{lme4} in choosing a parametrisation based on the Cholesky decomposition of $V(\nu)$ to define $\nu$ and using Powell's box-constrained quadratic optimiser {\tt bobyqa} \citep{bobyqa} to minimise the profile deviance under non-negativity constraints without requiring analytic derivatives.

\section{Examples}
\label{sec:eg}
\subsection{PISA education survey}
The 2012 edition of the OECD Programme for International Student Assessment (PISA) surveyed students and staff at schools in 65 countries \citep{pisa2012}. PISA provides school-level and student-level weights in its public-use datasets. We will fit a model to some of the data on mathematics attainment from New Zealand. These data are in the {\sf svylme} package as dataset {\tt nzmaths}. We have data on 4291 students at 177 schools.  We will model the mathematics attainment score. It is given as 5 `plausible values' sampled from a posterior distribution reflecting measurement error; we will use the first plausible value.  Our predictors are student gender, proportion of girls at the school, student/teacher ratio in mathematics, and two attitude scores. The scores measure openness to problem solving and mathematics self-efficacy.  The school gender proportion is close to 0, 0.5, or 1 for nearly all schools; we center it at 0.5.   There is only one PSU in stratum NZL0102; we combine it with stratum NZL0202, which is schools in the same medium size range.
Since the sampling units and model clusters are the same in this survey we can compare the results of the two pairwise estimators to the stagewise pseudolikelihood estimator implemented in Stata \citep{stata, rabe-hesketh-survey}, using the \citet{gk-scaling} scaling of weights. 

The model outputs are shown in Table~\ref{pisa-fit}. The model has student gender, proportion of girls at the school, their interaction, student/teacher ratio in mathematics, and the two attitude scores. 
The interpretation of the individual/school gender interaction may not be immediately obvious: it says that both boys and girls had higher average scores at single-sex than coeducational schools. We should be cautious about interpreting this causally, as single-sex schools in New Zealand differ from coeducational schools in other ways as well. 

The interpretations of the coefficients are qualitatively similar for the two estimators, and the estimated standard errors are also similar.  

\begin{table}
\caption{Analysis of mathematics achievement score from PISA 2012 in New Zealand schools, comparing pairwise (in R) and stagewise (in Stata) weighted likelihoods. $\dag$ indicates school-level variables. `Coef' is coefficient, `SE' is sandwich standard error, `JK' is jackknife standard error}
\label{pisa-fit}
\centering
\begin{tabular}{lrrrrrr|rr}
\toprule
& \multicolumn{3}{c}{Corr.\ pairs} & \multicolumn{3}{c}{All pairs} &\multicolumn{2}{c}{Stagewise}\\ 
& Coef. & SE & JK & Coef. & SE& JK &Coef. & SE   \\
\midrule
Intercept & 517.9 &12.6 & 13.6 &501.2&14.2&14.9& 496.6&15.3 \\ 
Male & 4.9 & 15.5 & 18.2&-1.0&15.9&17.0& 1.5 &14.4\\
\dag Proportion Girls$-0.5$ & 54.3 & 16.2 & 17.5&61.8&15.0&15.6&59.8&16.1\\
\dag Staff/student ratio & 0.0 &0.1& 0.1 &0.1&0.1&0.1& 0.1& 0.1\\
Male:\dag Prop Girls & -129.0 & 32.7 &36.5&-111.2&28.0&29.1&-96.7&28.4\\
Male:\dag staff ratio & -0.1 & 0.1 & 0.1&0.0&0.1&0.1&0.0 & 0.1\\
Math self-efficacy& 46.5 & 2.0 &2.0 &47.3&2.5&2.6&40.5&2.3\\
Problem-solving & 14.0 & 2.6 & 2.7&13.5&2.4&2.5&16.7&2.2 \\
\midrule
\multicolumn{6}{l}{Variance components (standard deviation scale)}\\
Intercept & 22.9 & --- & 5.8&25.7&---&5.8&28.3&8.3\\
Male& 10.9 &--- & 6.8&10.7&---&6.8&9.1 &5.4\\
Residual& 69.5 & --- & 1.4&70.9&---&1.3&70.0&1.2\\
\bottomrule
\end{tabular}
\end{table}

\subsection{Body Mass Index of Twins}
\label{eg:twins}

This example is taken from a vignette in the {\sf mets} R package \citep{mets-pkg}. The data consist of self-reports of body mass index by 11,188 individuals from  6917 same-sex twin pairs \citep{twins}. Approximately two-thirds of the twin pairs are monozygotic (`identical') and the other one-third dizygotic (`fraternal'). 

In this example we fit models with (a) a twin random-effect, (b) separate environmental and additive genetic effects, and (c) separate environment, additive genetic, and dominant genetic effects.  In table~\ref{twin-table} we compare the complete-data results from {\sf lme4} \citep{lme4}, {\sf lme4qtl} \citep{lme4qtl}, and our implementation, and compare estimates using all pairs and just the correlated pairs. To obtain complete-data results we set all the pairwise sampling probabilities to unity. In section~\ref{twin-sims} we also conduct simulations treating this data set as a population for sampling. The random-effect structures we consider all involve within-pair correlations. In the pure environmental model, the within-pair correlation is the same for all twin pairs and we write
$$\Xi = I + \tau_e I_{\text{pair}}$$
where the $(i,j)$ element of $I_{\text{pair}}$ indicates whether the two observations are in the same pair.  The two genetic structures alter the off-diagonal terms of the variance matrix. A dizygotic pair shares each allele at a locus identical-by-descent with their sibling with probability 1/2 and shares both alleles with probability 1/4.  They thus have half the additive genetic correlation of a monozygotic pair, and $1/4$ the dominant genetic correlation. The off-diagonal within-pair terms of the matrix $I_{\text{add}}$ are 1 for monozygotic  and 1/2 for dizygotic pairs; for  the matrix $I_{\text{dom}}$ they are 1 for monozygotic  and 1/4 for dizygotic pairs.  The two genetic models we consider are
\begin{align*}
\Xi &= I + \tau_e^2 I_{\text{pair}}+\tau_a^2I_{\text{add}}\\
\Xi &= I + \tau_e^2 I_{\text{pair}}+\tau_a^2I_{\text{add}}+\tau_d^2I_{\text{dom}}
\end{align*}

\begin{table}
\caption{Comparison of pairwise likelihood estimators and maximum likelihood estimators for genetic models in twin data. Maximum likelihood estimation used {\sf lme4} for the environment-only model and otherwise {\sf lme4qtl}.}
\label{twin-table}
\centering
\begin{tabular}{lrrrrrrrrr}
\toprule
 &&\multicolumn{2}{c}{Pairs}&&\multicolumn{2}{c}{Pairs}&&\multicolumn{2}{c}{Pairs}\\
Model &  ML & corr. & all &  ML & corr. & all &  ML & corr. & all \\
\midrule
Intercept & 18.68 & 18.57 & 18.66 & 18.68 & 18.56 & 18.66 & 18.68 & 18.57 & 18.66\\
Male& 0.12 & 0.12 & 0.12 & 0.12 & 0.12 & 0.12 & 0.12 & 0.12 & 0.12\\
Age (y)& 1.41 & 1.38 & 1.41 & 1.41 & 1.38 & 1.41 & 1.41 & 1.38 & 1.41\\
\midrule
$\tau_e$ & 2.18 & 2.17 & 2.18 & 1.81 & 1.80 & 1.80& 1.35 & 1.33 & 1.35\\
$\tau_a$ &---&---&---&1.23 & 1.22 & 1.22 & 1.06 & 1.05 & 1.06\\
$\tau_d$ &---&---&---&---&---&---&1.35& 1.37 & 1.35\\
$\sigma$ &2.60 & 2.60 & 2.60 & 2.60 & 2.60 & 2.60 & 2.60 & 2.60 & 2.60\\
\bottomrule
\end{tabular}
\end{table}

The three estimation approaches give very similar results for all three models; the nearly balanced design, with two observations for most twin pairs, will have reduced any difference.   

In this example, using all pairs by adjustment to the marginal likelihood took approximately 4 seconds for the environment-only model; using all pairs by direct computation took approximately 920 seconds (R 4.2.1, Apple M1).

\subsection{Milk yield in dairy cows}
\label{eg:milk}

\citet{pedigreemm-note} describe 3397 observations of milk yield from 1339 Holstein cows, which are correlated because they are in 57 herds and have only 38 different sires.   This is a much more extreme version of the non-nested genetic and environmental correlation that motivated our research.  In section~\ref{milk-sims} we conduct simulations based on these data; here we fit the model described by \citet{pedigreemm-note}  to the complete data using both maximum likelihood \citep{lme4qtl} and pairwise likelihood.  The model has two fixed-effect predictors: the lactation number for the cow (in the range 1--5) and the logarithm of the number of days in milk for the current lactation. There is also a random effect for herd, and a random effect for genetic relatedness. Writing $I_{\text{herd}}$ for the herd indicator matrix and $\Phi_{\text{gene}}$ for the genetic relatedness matrix:
$$\Xi = I+\tau_{\text{herd}}I_{\text{herd}}+ \tau_{\text{gene}}\Phi_{\text{gene}}$$

\begin{table}
\caption{Maximum likelihood and pairwise likelihood estimates for a genetic and environmental mixed model of milk yield in Holstein cows. The upper half of the table is fitted to the observed data; the lower half is fitted to data simulated from a maximum likelihood fit}
\label{milk-table}
\centering
\begin{tabular}{lrrrrrr}
\toprule
Estimator & Intercept & Lactation no. & log days & $\tau_{\text{herd}}$ & $\tau_{\text{gene}}$ & $\sigma$\\
\midrule
\multicolumn{7}{l}{Observed data}\\
MLE & 1.7 & $-0.11$ & 0.74 & 0.53 & 0.46 & 0.70\\
corr.\ pairs & 0.9 & $-0.05$ & 0.85 & 0.27 & 0.40 & 0.82\\
all pairs & 1.0 & $-0.05$ & 0.83 & 0.00 & 0.41 & 0.87\\
\midrule
\multicolumn{7}{l}{Simulated data}\\
MLE & 1.1 & $-0.09$ & 0.84 & 0.67 & 0.66& 0.82\\
corr.\ pairs & 1.0 & $-0.10$ & 0.81 & 0.65 & 0.67 & 0.85\\
all pairs & 1.1 & $-0.11$ & 0.81 & 0.58 & 0.66 & 0.88\\
\bottomrule
\end{tabular}

\end{table}

The upper half of table~\ref{milk-table} shows there is disagreement between the pairwise likelihood and maximum likelihood estimators.  Part of the reason is model misspecification. In particular, the distribution of between-herd variability in milk yield has longer tails than the assumed Normal distribution.  The pairwise likelihood estimators give more weight to large herds than the maximum likelihood estimator; the three estimators are not consistent for the same `least false' parameters.   We can examine the extent to which  model misspecification is the explanation, by repeating the estimation with milk yield data simulated from the maximum-likelihood model fit. Results are given in the lower half of table~\ref{milk-table}, and show much better agreement.  

The difference in population parameters is important when considering design-based estimation; in the presence of model misspecification a weighted pairwise likelihood estimator can only hope to be design-consistent for the pairwise-likelihood population parameter, not the population MLE.

\section{Simulations}
\label{sec:verify}
\label{sims}
We present three sets of simulations. The first examines the impact of using all pairs vs correlated pairs in a setting where many individuals have no correlated pairs.  The second simulates a setting where sampling units overlap with model clusters to varying degrees. Finally, we consider a probability-proportional-to-size sample from the Holstein cow data above, to illustrate that the proposed method is not limited to multistage stratified sampling.

\subsection{Twin simulations}
\label{twin-sims}

We conducted two simulations using the twin data from section~\ref{eg:twins} as the population. First, we oversampled twin pairs with large differences in BMI to demonstrate that the weighted pairwise likelihood estimator was approximately unbiased and to examine its loss of efficiency in comparison to naive maximum likelihood. The difference in BMI was divided into strata at the 40th, 60th and 80th percentiles, with 50 twin pairs taken from the first two strata, 150 from the third, and 400 from the fourth. 

Second, we subsampled individuals from the first sample independently with probability 1/2, to produce a sample where the half of observations are in no correlated pairs. We compare the correlated-pairs and all-pairs estimators.  In both cases we use a model with an environmental random effect and an additive genetic random effect, and with age and gender as fixed effects.

\begin{table}
\caption{Simulations from twin population. The upper half of the table compares naive ML and pairwise likelihood in a sample of twin pairs; the lower half compares all-pairs and correlated-pairs estimators in a subsample with 50\% singleton observations}
\label{tbl:twin-sim}
\centering
\begin{tabular}{lrrrrr|rrrrrr}
\toprule
& \multicolumn{5}{c}{Environment only} & \multicolumn{6}{c}{plus additive genetic}\\
&Int. &  Male & Age& $\tau_e$ &  $\sigma$ & Int. &  Male & Age & $\tau_e$ & $\tau_a$ &$\sigma$ \\
\midrule
Naive MLE& 19.8 & 0.11 & 1.2 & 1.5 &3.5 & 19.8 & 0.10 & 1.2 &1.5& 0.6 & 3.5\\
(SE) & 0.6 & 0.01 & 0.2 & 0.15 & 0.08& 0.6 & 0.01 & 0.2 & 0.3 & 0.6 & 0.08 \\
Corr. pairs&18.5& 0.12 & 1.4 & 2.1 & 2.6 & 18.5 &0.12 &1.4 & 1.8 & 1.0 & 2.6\\
(SE) &0.8 & 0.02 & 0.3 & 0.15 & 0.06 &0.8 & 0.02 & 0.3 & 0.4 & 0.7 & 0.06\\
\midrule
Corr. pairs & 18.5 & 0.12 & 1.4 & 2.1 & 2.6 & 18.5 &0.12 &1.4 & 1.4 &1.5 & 2.6\\
(SE) & 1.8 & 0.04 & 0.6 & 0.32 & 0.16 & 1.8 & 0.04 & 0.61 & 0.33 & 0.30 & 0.16\\
All pairs & 18.7 &0.12 & 1.4 & 2.1 & 2.6 & 18.6 & 0.12 &1.4 &1.5 &1.6 & 2.6\\
(SE) & 1.3 & 0.03 & 0.4 & 0.28 & 0.16 &1.3 & 0.03 & 0.43 & 0.34 & 0.28 & 0.16\\
\bottomrule
\end{tabular}

\end{table}

Table~\ref{tbl:twin-sim} shows the results. The first set of simulations shows relatively little difference in variability between the pairwise and naive ML estimators.  This is expected: pairwise likelihood would be maximum likelihood if all observations were pairs, so any additional variability is due to the weights, not to the estimation approach. Oversampling pairs with large differences in BMI results leads the naive ML estimator to overestimate the residual variance and underestimate the two random-effects variance components. The pairwise estimator remains approximately unbiased.

The second set of simulations confirms that the all-pairs estimator is more efficient than the correlated-pairs estimator for the fixed effects when many observations are singletons. There is no gain in efficiency for the random effects variances, where the singletons do not contribute any information. Interestingly, there is also no gain in efficiency for the residual variance, perhaps because it is not separately identifiable in singletons.

\subsection{Partially crossed effects simulations}

Here we simulate random effects that do not nest with the survey design, under a strongly informative design.  We compare the pairwise likelihood estimator to naive maximum likelihood ignoring the sampling and to a design-weighted linear regression that does not estimate the variance components. 

We begin with a population on a square $400\times 400$ grid. The columns of the grid are the primary sampling units of the design.  The model clusters are controlled by a parameter $N_{\text{overlap}}$. For the first $N_{\text{overlap}}$ rows, model cluster $i$ is in column $i$ and so overlaps PSU $i$. In row $i+k$, the observation from cluster $i$ is in column $i+k$ and so in PSU $i+k$; the addition is modulo 400 so the clusters wrap around at the edge of the population grid.  We generate random intercepts $u_i$ for cluster by sampling random intercepts from $N(0,\tau^2)$ and then sorting into increasing order, residuals $N(0,\sigma^2)$. There are two covariates: $Z$ is iid $N(0,1)$ and $X$ is the column number modulo 40.  When sampling, we take sets of 40 contiguous columns as a stratum, giving ten strata, and take stratified random sample of (20, 5, 4, 3, 2, 2, 3, 4, 5, 20) PSUs respectively from the strata. At stage two we take  20 elements in the first and last PSUs sampled and 8 from each of the other 66 PSUs sampled, giving a total of 568. For each sampled population, bias is estimated with the median of the simulation results and the simulation standard error is estimated by the scaled median absolute deviation; these are then averaged over simulated finite populations. Jackknife standard errors use a stratified `JKn' cluster jackknife. The simulation code is in the Supplemental Materials. 

Table~\ref{one-crossed} shows the results of this simulation.  The pairwise likelihood estimates are approximately unbiased in all settings.  The naive ML estimates are severely biased for the random effects standard deviation $\tau_0$, as was intended for the sampling design, and show some bias for the fixed intercept but are otherwise approximately unbiased.  The simulation standard errors show that pairwise likelihood estimator is substantially less efficient than the ML estimator, especially when the overlap is smaller.  The loss of efficiency in the fixed effects is due partly to variation in the weights and partly to uncertainty in the variance components, as is indicated by the intermediate loss of efficiency for a simple design-weighted least squares estimator.   In a familiar phenomenon for both survey estimators and linear mixed models, the sandwich variance estimator underestimates the simulation standard errors to some degree; the jackknife overestimates them.

\begin{table}
\caption{One random effect partially crossed with the design. Summary of 1000 simulations for each of 100 finite populations under each condition. Naive ML uses {\sf lme4}. }
\label{one-crossed}
\centering
\begin{tabular}{llrrrrr}
\toprule
&& $\beta_0=0$ & $\beta_x=1$ & $\beta_z=1$ & $\tau^2=1$ &  $\sigma^2=1$\\
\midrule
Overlap = 25\%\\
Naive ML & Bias &  -0.13 & 0.003 & 0.00 & 0.11 & -0.025 \\
& Sim. SE & 0.10 & 0.005 & 0.05  &0.11 &0.08 \\
& Model $\widehat{SE}$ &  0.12 & 0.005 & 0.05& --- & ---\\
Pairwise & Bias  & -0.03 & 0.001 & 0.00 & -0.05 & -0.03 \\
& Sim SE & 0.25 & 0.010 & 0.10 & 0.26 & 0.15\\
& Sandwich $\widehat{SE}$ & 0.25 & 0.008 & 0.093& --- & --- \\
& Jackknife $\widehat{SE}$ & 0.28 & 0.010 & 0.116 & 0.28 & 0.17\\
Least squares & Bias & -0.05 & 0.003 & 0.000& --- & ---\\\
& Sim SE & 0.16 & 0.007 & 0.08 & --- & ---\\\
& Sandwich $\widehat{SE}$ & 0.15 & 0.007 & 0.08 & --- & ---\\
\midrule
Overlap = 75\%\\
Naive ML & Bias &  -0.27 & 0.004 & 0.00 & 0.72 & 0.05 \\
& Sim. SE & 0.16 & 0.007 & 0.05  &0.13 &0.07 \\
& Model $\widehat{SE}$ &  0.18 & 0.007 & 0.05& --- & ---\\
Pairwise & Bias  & -0.12 & 0.006 & 0.00 & -0.02 & -0.04 \\
& Sim SE & 0.28 & 0.013 & 0.09 & 0.19 & 0.13\\
& Sandwich $\widehat{SE}$ & 0.25 & 0.010 & 0.08& --- & --- \\
& Jackknife $\widehat{SE}$ & 0.31 & 0.014 & 0.11 & 0.23 & 0.15\\
Least squares & Bias & -0.16 & 0.008 & 0.00& --- & ---\\\
& Sim SE & 0.20 & 0.010 & 0.07 & --- & ---\\\
& Sandwich $\widehat{SE}$ & 0.19 & 0.009 & 0.07 & --- & ---\\
\bottomrule
\end{tabular}
\end{table}

\subsection{Gene/environment simulations}
\label{milk-sims}
In this example we use a relatively complex  design to subsample from the Holstein cow database analysed in section~\ref{eg:milk} above.  We sample ten herds with probability proportional to the total milk yield, using Till\'e's algorithm \citep{tille-pps, pkg-sampling}, which provides pairwise sampling probabilities.  Sampling with probability proportional to size induces negative correlations between sampling indicators, and even for the same marginal probabilities these correlations will depend on the sampling algorithm.  As above, we sample both from the observed data and from simulated data based on a maximum likelihood fit to these data.  The sampling is only weakly informative.  The cluster (herd) sizes vary from 1 to 255, so it is not surprising that pairwise likelihood is relatively inefficient.  Interestingly, the loss of efficiency is not apparent for the genetic variance component. 

\begin{table}
\caption{Simulation median bias (compared to complete data) and standard error of maximum likelihood and pairwise likelihood estimates for samples from the Holstein milk yield population. The upper half of the table is fitted to the observed data; the lower half is fitted to data simulated from a maximum likelihood fit.}
\label{milk-sim-table}
\centering
\begin{tabular}{lrrrrrr}
\toprule
Estimator & Intercept & Lactation no. & log days & $\tau^2_{\text{herd}}$ & $\tau^2_{\text{gene}}$ & $\sigma^2$\\
\midrule

\multicolumn{7}{l}{Observed data}\\
Naive MLE & -0.18& -0.00 & 0.04 & -0.02 & -0.06 & 0.01\\
Sim SE & 0.56 & 0.03 & 0.08 & 0.04 & 0.08& 0.04 \\
corr.\ pairs & -0.04 & 0.00 &0.01 & 0.10 & 0.00 & -0.11\\
Sim SE & 1.16 & 0.05 & 0.18 & 0.13 & 0.08 & 0.03\\
all pairs & 0.10 & 0.00& -0.01 & 0.15 & -0.01 & -0.17\\
Sim SE & 1.18 & 0.05 & 0.19 & 0.13 & 0.08 & 0.12\\
\midrule
\multicolumn{7}{l}{Simulated data}\\
Naive MLE & -0.24 & 0.01 & 0.02 & 0.02 & -0.04 &-0.01\\
Sim SE & 0.49 & 0.02 & 0.08 & 0.03& 0.11 & 0.03\\
corr.\ pairs &-0.14 & 0.00 & 0.02 & -0.07 & -0.02 & 0.05\\
Sim SE& 1.14 & 0.05 &0.19 &0.11 &0.11 & 0.10\\
all pairs & -0.13 & 0.00 & 0.02 & -0.07 & -0.02 & 0.05\\
Sim SE& 1.03 & 0.05 & 0.16 & 0.11 &0.12 & 0.10\\
\bottomrule
\end{tabular}

\end{table}

\section{Discussion}
\label{sec:conc}

Simulations here and in \citet{svylme-stat} confirm that weighted pairwise likelihood is effectively design-unbiased even under strongly informative sampling, but at a cost in efficiency, especially when sample clusters are of very different sizes. As a consequence, we recommend weighted pairwise likelihood estimation  when sampling is expected to be informative and the variance components are of substantive interest, design variables are either non-available or not appropriate for inclusion in the model (eg, with outcome-dependent sampling). If sampling is non-informative or can be made non-informative by adjusting for design variables, naive maximum likelihood may be preferable, and if the variance components are not of interest, fitting a design-weighted linear model will give greater precision. The examples in \citet{svylme-stat} indicate that the loss in precision for fixed effects is small when sample clusters are all the same size.

The loss of efficiency is somewhat surprising. Statistical folklore says that the pairwise likelihood estimator has good efficiency for mixed models. For example, \citet{composite-review} says  ``\dots most simulation studies show that some version of composite likelihood has high efficiency'' before going on to note one exception. This conclusion has been drawn largely from models for binary and count data data. It appears that there is non-negligible information loss  linear mixed models, especially for the variance components $\nu$. \citet{chen-thesis} investigated reweighting the pairwise loglikelihood to increase efficiency, but did not find any meaningful gains in this context.

Inference using pairwise likelihood is valid very generally, including to generalised linear mixed models. Computation, however, is  more challenging. 
Each pairwise likelihood contribution involves only two observations and so does not give good estimates of the realised random effects to centre the integration. When there is more than one variance parameter $\nu$, the standard equations for the pairwise BLUPs will be singular. Further research is needed on pooling information across pairs to estimate the realised random effects without breaking the pairwise reweighting. \citet{yi-rao-mixed} used ordinary Gauss--Hermite quadrature, which is feasible when the model clusters and sampling units are the same and the number of variance components is small. It is not feasible for general models and designs. When the sampling units and model clusters are not nested, or when there is correlation between random effects in different sampling units, the likelihood even for a single pair of observations and a single variance component may involve high-dimensional integrals. 

Even REML estimation presents some challenges for a general implementation.  The benefit of REML is correct accounting for degrees of freedom used up in estimating the fixed effects, and REML is of most value when the number of predictors $p$ is not small compared to the number of observations $n$.  When that is the case, it does not seem reasonable to assume the population model would have only the same set of predictors and the same $p$, and even if it did, $N-p$ for the population $N$ would be much larger than $p$ making the REML criterion very similar to ML. 

Bayesian estimation has led to better statistical performance in unweighted estimation for generalised linear mixed models (at some computational cost) and so is of interest for design-weighted estimation even to frequentists.  \citet{savitsky-pairwise} presented a pseudo-Bayesian composite likelihood approach and the same authors have also derived pseudo-Bayesian full likelihood estimators \citep{savitsky-bayesian}. Comparisons with these would be of interest for future research. 

The pairwise loglikelihood requires pairwise weights for its definition. Under multistage cluster sampling these weights can be computed straightforwardly if the sampling probabilities at each stage are known.  For example, pairwise weights for HCHS/SoL are described by \citet{hchs-gee}.  The ideal way to apply post-stratification and raking adjustments is not clear: should these be recomputed directly for pairs or can pairwise weights be computed from separately raked marginal weights?

Many surveys, however, do not provide stage-specific weights in their public-use data. We conjecture that consistent estimation is not possible in general with just a single overall weight for all the stages of sampling, but it may be possible to achieve acceptable estimation in practice. \citet{savitsky-bayesian} argue that a single overall weight is sufficient when distinct random effects are independent. There are various approximations to pairwise sampling probabilities that have been used to define standard error estimates in the survey literature;  further research is needed into which of these approximations are useful in constructing pairwise weights. 

\bigskip
\begin{center}
{\large\bf SUPPLEMENTAL MATERIALS}
\end{center}

\begin{description}

\item[R-package] R-package `svylme' containing code to perform the methods described in the article: \url{https://github.com/tslumley/svy2lme}

\item[Simulation code] R scripts reproducing the simulation results: in the package, in the {\tt inst/scripts} directory

\item[Example code] R scripts reproducing the illustrative examples: in the package, in the {\tt inst/scripts} directory

\item[Asymptotics] Outline proof of consistency and asymptotic normality of the estimators for both linear and generalised linear mixed models.

\end{description}

\bibliography{svylme}

\begin{thebibliography}{48}
\providecommand{\natexlab}[1]{#1}
\providecommand{\url}[1]{\texttt{#1}}
\expandafter\ifx\csname urlstyle\endcsname\relax
  \providecommand{\doi}[1]{doi: #1}\else
  \providecommand{\doi}{doi: \begingroup \urlstyle{rm}\Url}\fi

\bibitem[Baldi and Rinott(1989)]{baldi-rinott}
P.~Baldi and Y.~Rinott.
\newblock On normal approximations of distributions in terms of dependency
  graphs.
\newblock \emph{Annals of Probability}, 17:\penalty0 1646--1650, 1989.

\bibitem[Bates et~al.(2015)Bates, M\"{a}chler, Bolker, and Walker]{lme4}
D.~Bates, M.~M\"{a}chler, B.~Bolker, and S.~Walker.
\newblock Fitting linear mixed-effects models using lme4.
\newblock \emph{Journal of Statistical Software}, 67\penalty0 (1):\penalty0
  1--48, 2015.
\newblock ISSN 1548-7660.

\bibitem[Bolthausen(1982)]{bolthausen-stein}
E.~Bolthausen.
\newblock On the central limit theorem for stationary mixing random fields.
\newblock \emph{The Annals of Probability}, 10\penalty0 (4):\penalty0
  1047--1050, 1982.

\bibitem[Bradley(2005)]{bradley-mixing}
R.~C. Bradley.
\newblock {Basic Properties of Strong Mixing Conditions. A Survey and Some Open
  Questions}.
\newblock \emph{Probability Surveys}, 2\penalty0 (none):\penalty0 107 -- 144,
  2005.

\bibitem[Brewer(2002)]{brewer}
K.~Brewer.
\newblock \emph{Combined Survey Sampling Inference: Weighing Basu's Elephants}.
\newblock Hodder Education, London, UK, 2002.

\bibitem[Charlton et~al.(2022)Charlton, Rasbash, Browne, Healy, and
  Cameron]{mlwin}
C.~Charlton, J.~Rasbash, W.~Browne, M.~Healy, and B.~Cameron.
\newblock \emph{MLwiN}.
\newblock Centre for Multilevel Modelling, University of Bristol, 3.06 edition,
  2022.

\bibitem[Chen(2021)]{chen-thesis}
S.~D. Chen.
\newblock Investigation into the efficiency of design-weighted pairwise
  log-likelihoods for linear mixed model estimation.
\newblock Master's thesis, University of Auckland, Auckland, New Zealand.,
  2021.
\newblock URL \url{https://researchspace.auckland.ac.nz/handle/2292/56126}.

\bibitem[Graubard and Korn(1996)]{gk-scaling}
B.~I. Graubard and E.~L. Korn.
\newblock Modelling the sampling design in the analysis of health surveys.
\newblock \emph{Statistical Methods in Medical Research}, 5\penalty0
  (3):\penalty0 263--281, 1996.

\bibitem[Guyon(1995)]{guyon}
X.~Guyon.
\newblock \emph{Random Fields on a Network: Modeling, Statistics, and
  Applications}.
\newblock Springer-Verlag, 1995.

\bibitem[Heagerty and Lele(1998)]{heagerty-lele}
P.~J. Heagerty and S.~R. Lele.
\newblock A composite likelihood approach to binary spatial data.
\newblock \emph{Journal of the American Statistical Association}, 93\penalty0
  (443):\penalty0 1099--1111, 1998.

\bibitem[Holst and Scheike(2023)]{mets-pkg}
K.~K. Holst and T.~Scheike.
\newblock mets: Analysis of multivariate event times, 2023.
\newblock URL \url{https://CRAN.R-project.org/package=mets}.
\newblock R package version 1.3.2.

\bibitem[Huang(2019)]{huang}
X.~Huang.
\newblock \emph{Mixed Models for Complex Survey Data}.
\newblock PhD thesis, University of Auckland, Auckland, New Zealand, March
  2019.

\bibitem[Janson(1988)]{janson-clt}
S.~Janson.
\newblock Normal convergence by higher semiinvariants with applications to sums
  of dependent random variables and random graphs.
\newblock \emph{The Annals of Probability}, 16\penalty0 (1):\penalty0 305 --
  312, 1988.

\bibitem[Korkeila et~al.(1991)Korkeila, Kaprio, Rissanen, and Koskenvuo]{twins}
M.~Korkeila, J.~Kaprio, A.~Rissanen, and M.~Koskenvuo.
\newblock Effects of gender and age on the heritability of body mass index.
\newblock \emph{International journal of obesity}, 15\penalty0 (10):\penalty0
  647—654, October 1991.

\bibitem[Laird and Ware(1982)]{laird-ware-mixed}
N.~M. Laird and J.~H. Ware.
\newblock Random-effects models for longitudinal data.
\newblock \emph{Biometrics}, 38:\penalty0 963--74, 1982.

\bibitem[Lavange et~al.(2010)Lavange, Kalsbeek, Sorlie, Avilés-Santa, Kaplan,
  Barnhart, Liu, Giachello, Lee, Ryan, Criqui, and Elder]{hchs-design}
L.~Lavange, W.~Kalsbeek, P.~Sorlie, L.~Avilés-Santa, R.~Kaplan, J.~Barnhart,
  K.~Liu, A.~Giachello, D.~Lee, J.~Ryan, M.~Criqui, and J.~Elder.
\newblock Sample design and cohort selection in the {Hispanic Community Health
  Study/Study of Latinos.}
\newblock \emph{Annals of Epidemiology}, 20\penalty0 (8):\penalty0 642--9,
  2010.

\bibitem[Lin et~al.(2014)Lin, Tao, Kalsbeek, Zeng, Gonzalez, Fernández-Rhodes,
  Graff, Koch, North, and Heiss]{hchs-gee}
D.~Lin, R.~Tao, W.~Kalsbeek, D.~Zeng, F.~Gonzalez, 2nd, L.~Fernández-Rhodes,
  M.~Graff, G.~Koch, K.~North, and G.~Heiss.
\newblock Genetic association analysis under complex survey sampling: the
  {Hispanic Community Health Study/Study of Latinos}.
\newblock \emph{American Journal of Human Genetics}, 95\penalty0 (6):\penalty0
  675--88, 2014.

\bibitem[Lindsay(1988)]{lindsay-composite}
B.~G. Lindsay.
\newblock Composite likelihood methods.
\newblock \emph{Contemporary Mathematics}, 80:\penalty0 221--239, 1988.

\bibitem[Liu(2007)]{liu-pairwise-thesis}
J.~Liu.
\newblock \emph{Multivariate Ordinal Data Analysis with Pairwise Likelihood and
  Its Extension to SEM}.
\newblock PhD thesis, University of California Los Angeles, 2007.

\bibitem[Lumley(2004)]{survey-paper}
T.~Lumley.
\newblock Analysis of complex survey samples.
\newblock \emph{Journal of Statistical Software}, 9\penalty0 (1):\penalty0
  1--19, 2004.
\newblock R package verson 2.2.

\bibitem[Lumley(2023{\natexlab{a}})]{survey-pkg}
T.~Lumley.
\newblock survey: analysis of complex survey samples, 2023{\natexlab{a}}.
\newblock URL \url{https://CRAN.R-project.org/package=survey}.
\newblock R package version 4.2.

\bibitem[Lumley(2023{\natexlab{b}})]{svylme-pkg}
T.~Lumley.
\newblock svylme: Linear mixed models for complex survey data,
  2023{\natexlab{b}}.
\newblock URL \url{https://CRAN.R-project.org/package=svylme}.
\newblock R package version 1.3.

\bibitem[Lumley and Huang(2023)]{svylme-stat}
T.~Lumley and X.~Huang.
\newblock Linear mixed models for complex survey data: implementing and
  evaluating pairwise likelihood.
\newblock \emph{Stat}, 2023.

\bibitem[Muth\'en and Muth\'en(2012)]{mplus}
L.~K. Muth\'en and B.~O. Muth\'en.
\newblock \emph{Mplus User's Guide}.
\newblock Muth\'en \& Muth\'en, Los Angeles, CA, seventh edition, 2012.

\bibitem[OECD(2013)]{pisa2012}
OECD.
\newblock \emph{{PISA 2012 Assessment and Analytical Framework: Mathematics,
  Reading, Science, Problem Solving and Financial Literacy}}.
\newblock OECD Publishing, 2013.

\bibitem[Pfeffermann et~al.(1998)Pfeffermann, Skinner, Holmes, Goldstein, and
  Rasbash]{pfeffermann-mixed}
D.~Pfeffermann, C.~J. Skinner, D.~J. Holmes, H.~Goldstein, and J.~Rasbash.
\newblock Weighting for unequal selection probabilities in multilevel models.
\newblock \emph{Journal of the Royal Statistical Society, Series B},
  60:\penalty0 23--40, 1998.

\bibitem[Powell(2009)]{bobyqa}
M.~J.~D. Powell.
\newblock The {BOBYQA} algorithm for bound constrained optimization without
  derivatives.
\newblock Technical Report DAMTP 2009/NA06, Department of Applied Mathematics
  and Theoretical Physics, Cambridge University, 2009.

\bibitem[{R Core Team}(2023)]{r-itself}
{R Core Team}.
\newblock \emph{R: A Language and Environment for Statistical Computing}.
\newblock R Foundation for Statistical Computing, Vienna, Austria, 2023.
\newblock URL \url{https://www.R-project.org/}.

\bibitem[Rabe-Hesketh and Skrondal(2006)]{rabe-hesketh-survey}
S.~Rabe-Hesketh and A.~Skrondal.
\newblock Multilevel modelling of complex survey data.
\newblock \emph{Journal of the Royal Statistical Society, Series A},
  169:\penalty0 805--827, 2006.

\bibitem[Rabe-Hesketh et~al.(2004)Rabe-Hesketh, Skrondal, and
  Pickles]{gllamm-framework}
S.~Rabe-Hesketh, A.~Skrondal, and A.~Pickles.
\newblock Generalized multilevel structural equation modelling.
\newblock \emph{Psychometrika}, 69\penalty0 (2):\penalty0 167--190, 2004.

\bibitem[Rao and Wu(1988)]{raowu-boot}
J.~N.~K. Rao and C.~F.~J. Wu.
\newblock Resampling inference with complex survey data.
\newblock \emph{Journal of the American Statistical Association}, 83\penalty0
  (401):\penalty0 231--241, 1988.

\bibitem[Rao et~al.(2014)Rao, Verret, and Hidiroglou]{rao-mixed}
J.~N.~K. Rao, F.~Verret, and M.~A. Hidiroglou.
\newblock A weighted composite likelihood approach to inference for two-level
  models from survey data.
\newblock \emph{Survey Methodology}, 39:\penalty0 263--282, 2014.

\bibitem[Ross(2011)]{stein-ross}
N.~Ross.
\newblock {Fundamentals of Stein’s method}.
\newblock \emph{Probability Surveys}, 8\penalty0 (none):\penalty0 210 -- 293,
  2011.

\bibitem[Savitsky and Williams(2022)]{savitsky-bayesian}
T.~D. Savitsky and M.~R. Williams.
\newblock Pseudo {Bayesian} mixed models under informative sampling.
\newblock \emph{Journal of Official Statistics}, 38\penalty0 (3):\penalty0
  901--928, 2022.

\bibitem[Schneider(2023)]{svrep-pkg}
B.~Schneider.
\newblock svrep: Tools for creating, updating, and analyzing survey replicate
  weights, 2023.
\newblock URL \url{https://CRAN.R-project.org/package=svrep}.
\newblock R package version 0.5.1.

\bibitem[Shao(1996)]{shao-resample}
J.~Shao.
\newblock Resampling methods in sample surveys.
\newblock \emph{Statistics}, 27\penalty0 (3-4):\penalty0 203--237, 1996.

\bibitem[StataCorp(2023)]{stata}
StataCorp.
\newblock \emph{Stata Statistical Software: Release 18.}
\newblock StataCorp, College Station, TX, 2023.

\bibitem[Till\'e(1996)]{tille-pps}
Y.~Till\'e.
\newblock An elimination procedure for unequal probability sampling without
  replacement.
\newblock \emph{Biometrika}, 83\penalty0 (1):\penalty0 238--241, 1996.

\bibitem[Till\'e and Matei(2021)]{pkg-sampling}
Y.~Till\'e and A.~Matei.
\newblock \emph{sampling: Survey Sampling}, 2021.
\newblock URL \url{https://CRAN.R-project.org/package=sampling}.
\newblock R package version 2.9.

\bibitem[Vaart(1998)]{vdv-book}
A.~W. v.~d. Vaart.
\newblock \emph{Asymptotic Statistics}.
\newblock Cambridge Series in Statistical and Probabilistic Mathematics.
  Cambridge University Press, 1998.

\bibitem[Varin and Vidoni(2006)]{varin-vidoni-pairwise}
C.~Varin and P.~Vidoni.
\newblock Pairwise likelihood inference for ordinal categorical time series.
\newblock \emph{Computational Statistics \& Data Analysis}, 51\penalty0
  (4):\penalty0 2365--2373, 2006.
\newblock Nonlinear Modelling and Financial Econometrics.

\bibitem[Varin et~al.(2011)Varin, Reid, and Firth]{composite-review}
C.~Varin, N.~Reid, and D.~Firth.
\newblock An overview of composite likelihood methods.
\newblock \emph{Statistica Sinica}, 21\penalty0 (1):\penalty0 5--42, 2011.

\bibitem[Vasdekis et~al.(2012)Vasdekis, Cagnone, and
  Moustaki]{vasdekis-ordinal}
V.~G.~S. Vasdekis, S.~Cagnone, and I.~Moustaki.
\newblock A composite likelihood inference in latent variable models for
  ordinal longitudinal responses.
\newblock \emph{Psychometrika}, 77\penalty0 (3):\penalty0 425--441, 2012.

\bibitem[Vasdekis et~al.(2014)Vasdekis, Rizopoulos, and
  Moustaki]{vasdekis-pairwise}
V.~G.~S. Vasdekis, D.~Rizopoulos, and I.~Moustaki.
\newblock {Weighted pairwise likelihood estimation for a general class of
  random effects models}.
\newblock \emph{Biostatistics}, 15\penalty0 (4):\penalty0 677--689, 05 2014.

\bibitem[Vazquez et~al.(2010)Vazquez, Bates, Rosa, Gianola, and
  Weigel]{pedigreemm-note}
A.~I. Vazquez, D.~M. Bates, G.~J.~M. Rosa, D.~Gianola, and K.~A. Weigel.
\newblock {Technical note: An R package for fitting generalized linear mixed
  models in animal breeding}.
\newblock \emph{Journal of Animal Science}, 88\penalty0 (2):\penalty0 497--504,
  02 2010.

\bibitem[Williams and Savitsky(2018)]{savitsky-pairwise}
M.~R. Williams and T.~D. Savitsky.
\newblock {Bayesian pairwise estimation under dependent informative sampling}.
\newblock \emph{Electronic Journal of Statistics}, 12\penalty0 (1):\penalty0
  1631 -- 1661, 2018.

\bibitem[Yi et~al.(2016)Yi, Rao, and Li]{yi-rao-mixed}
G.~Y. Yi, J.~N.~K. Rao, and H.~Li.
\newblock A weighted composite likelihood approach for analysis of survey data
  under two-level models.
\newblock \emph{Statistica Sinica}, 26:\penalty0 569--587, 2016.

\bibitem[Ziyatdinov et~al.(2018)Ziyatdinov, V\'azquez-Santiago, Brunel,
  Martinez-Perez, Aschard, and Soria]{lme4qtl}
A.~Ziyatdinov, M.~V\'azquez-Santiago, H.~Brunel, A.~Martinez-Perez, H.~Aschard,
  and J.~M. Soria.
\newblock lme4qtl: linear mixed models with flexible covariance structure for
  genetic studies of related individuals.
\newblock \emph{BMC Bioinformatics}, 68\penalty0 (19):\penalty0 1--5, 2018.

\end{thebibliography}

\appendix

\section{Supplement: outline of asymptotics}

\subsection{Gaussian population pairwise likelihood}
Let $\tilde Y$ and $\tilde X$ be constructed by stacking all the outcome vectors and design matrices for pairs, so that with $|{\cal P}|$ pairs they have $2|\cal P|$ rows. Write $\sigma^2\tilde \Xi(\nu)$ for the $2|{\cal P}|\times 2|{\cal P}|$ matrix whose diagonal $2\times 2$ blocks are the modelled variance matrix of the corresponding pair of $\tilde Y$. The pairwise loglikelihood is 
$$\ell(Y;\beta,\nu,\sigma^2) = -\frac{1}{2}\log\left|\sigma^2\tilde \Xi \right|-\frac{1}{2}(\tilde Y -\tilde X\beta)^T\tilde \Xi^{-1}(\tilde Y -\tilde X\beta).$$
This is a quadratic form in Gaussian random variables, so it has the distribution of
$$\sum_{i=1}^N\lambda_i Z^2_i,$$
where $Z$ are independent Gaussian and $\lambda$ are the eigenvalues of $\sigma^{-2}\Xi^{-1}(\nu)\mathrm{var}[Y]$. By the Lindeberg central limit theorem this is asymptotically normal under conditions on $\max_i \lambda_i$ and the means of the $Z_i$. In particular, if the model is correctly specified, then at $\theta=\theta_0$ we have $\lambda_i=1$ and $Z_i\sim N(0,1)$, so $\ell$ is asymptotically Normal by the classical CLT:
$$N^{-1/2}\left(\ell-E[\ell]\right)\stackrel{d}{\to} N(0,\omega).$$
Standard smoothness arguments \citep[section 5.6]{vdv-book} now imply that $\hat\theta$ is asymptotically Normal as long as $\Xi(\nu_0)$ is non-singular, $\sigma^2_0$ is strictly positive, and $\nu_0$ is in the interior the parameter space.

\subsection{CLTs under graph-structured dependence}

A dependence graph $\Gamma_n$ for a set $\cal S$ of random variables $X_n$ is a graph whose vertices are the variables, such that two subsets ${\cal S}_A$, ${\cal S}_B$ of $\cal S$ are independent if no variable in ${\cal S}_A$ has an edge to a variable in  ${\cal S}_B$.  Alternatively, $\{{\cal N}_i\}$ is a set of dependence neighbourhoods for $\cal S$ if $S_A$ is independent of $S_B$ whenever $\bigcup_{i\in A} {\cal N_i}$ and $\bigcup_{i\in B} {\cal N_i}$ are disjoint.  Every dependence graph gives rise to a set of dependence neighbourhoods where ${\cal N}_i$ is the set of vertices adjacent to $i$. The idea of dependence graphs or neighbourhoods is to express \emph{sparse dependence}, where most pairs of small sets of variables are independent but the dependence cannot be represented just  by non-overlapping clusters.

This first result is due originally to \citet{baldi-rinott}. It uses Stein's method for normal approximation. In the form I quote it is from \citet{stein-ross}, a review of Stein's method in probability. 

\begin{theorem}[Ross, Theorem 3.6]
\label{steinclt}
Let $X_1,\dots,X_n$ be random variables such that $E[X_i^4]<\infty$, $E[X_i]=0$, $\sigma^2=\mathrm{var}[\sum_i X_i]$ and define $S=\sum_i X_i/\sigma$. Let the collection $\{X_i\}$ have dependency neighbourhoods ${\cal N}_i$ and define $M=\max_i|{\cal N}_i|$. Then for $Z$ a standard Normal variable
$$d_W(S,Z)\leq \frac{M^2}{\sigma^3}\sum_i E|X_i|^3+\frac{\sqrt{28}M^{3/2}}{\sqrt{\pi}\sigma^2}\sqrt{\sum_i E\left[X_i^4\right]}$$
where $d_W$ is the Wasserstein distance.
\end{theorem}

The second theorem is the first explicit use of dependence graphs, by \citet{janson-clt}. It has stronger tail assumptions and does not provide an explicit bound, but it has weaker assumptions on the marginal variance. The theorem was proved using convergence of the cumulants of the sum to the cumulants of a Normal distribution.  
\begin{theorem}[Janson, Theorem 2]
\label{jansonclt}
Suppose that for each $n$, $\{X_n\}_1^{N_n}$ is a family of bounded random variables, $|X_{ni}|\leq A_n$ a.s. Suppose further that $\Gamma_n$ is a dependency graph for this family and let $M_n$ be the maximal degree of $\Gamma_n$. Let $S_n=\sum_1^{N_n}$ and $\sigma^2_n=\mathrm{var}[S_n]$. If there exists an integer $r$ such that 
$$(N_n/M_n)^{1/r}M_nA_n/\sigma_n\to 0$$
then
$$(S-ES_n)/\sigma_n\stackrel{d}{\to} N(0,1).$$
\end{theorem}

If the dependence were generated by some sort of random effects (not necessarily Gaussian or additive), with non-zero variance components for each grouping factor, we would expect $\sigma^{2}_n=\Theta(Mn)$. If in addition the third and fourth moments of $X_i$ are uniformly bounded, the Wasserstein distance in theorem~\ref{steinclt} converges to zero whenever $M/n\to 0$. 

If the dependence is generated by sampling, it is not necessarily true that $\sigma^2_n$ grows faster than $n$. It is still reasonable to assume that $\sigma^2_n/n\not\to 0$, ie, that the design effect is bounded away from zero.  If the $X_i$ are uniformly bounded, theorem~\ref{jansonclt} applies directly as long as $M/N\to 0$. If the $X_i$ are not uniformly bounded but their variances are, a standard truncation argument also allows the use of theorem~\ref{jansonclt}

\subsection{Sampling}
The weighted pairwise loglikelihood is 
$$\hat\ell(\theta)=\sum_{i,j\in{\cal P}_N} \frac{R_{ij}}{\pi_{ij}}\ell_{ij}(\theta)$$
where ${\cal P}_N$ is the set of pairs in the population correlated under the model (for correlated-pairs estimation) or the set of all pairs in the population (for all-pairs estimation). 

We assume
(i) the sampling probabilities are bounded above and below, in the sense that there exist finite $a$ such that $0<a<\min_i\{\pi_{ij}\}/\max_i\{\pi_{ij}\}$ for all $n$;
(ii) the maximum number of pairs including any individual observation is $O(N^{1-\delta})$ for some $\delta>0$; and
(iii) The loglikelihood contributions $\ell_{ij}(\theta_0)$ have uniformly bounded variances under sampling from the model.
(iv) The design effect is bounded away from zero.

Assumption (iii) allows truncation of the loglikelihood contributions to be uniformly bounded for almost every sequence of populations. We may then take $A_n$ constant in Theorem~\ref{jansonclt}. By assumption $(iv)$, $\sigma_n$ is bounded below by a multiple of $\sqrt{n}$ and $M_n/N_n=O(N^{-\delta})$.

Theorem~\ref{jansonclt} then implies that 
$$a\hat\ell(\theta)= \sum_{i,j\in{\cal P}_N} \frac{R_{ij}a}{\pi_{ij}}\ell_{ij}(\theta)$$
is asymptotically Normal. Asymptotic normality of $\hat\theta$ then follows from standard smoothness arguments.

\subsection{Generalised linear mixed models under pairwise likelihood with uncorrelated random effects}

We consider the data-generating process for the population and the sampling separately. Both involve graph-structured dependence.  In the population, two pairs are dependent if an observation in one pair shares a random effect with an observation in the other pair.  Write $M_N$ for the maximal degree of the dependence graph in the population of size $N$. Theorem~\ref{steinclt} will apply if
(i) $M_N=O(N)$ and
(ii) $\theta$ is not on the boundary of the parameter space, implying $\sigma_N\neq o(M_N)$
(iii) the third and fourth moments of $\ell_{ij}(\theta_0)$ are uniformly bounded

Asymptotical normality of the census loglikelihood follows from theorem~\ref{steinclt}.  Under the same assumptions and by the same arguments as for the linear mixed model, asymptotic normality of $\hat\ell(\theta_0)$ for almost every sequence of populations follows from theorem~\ref{jansonclt}

\subsection{Correlated random effects}
When the population model contains correlated random effects, theorem~\ref{steinclt} is not sufficient, but there are a variety of extensions that can be used in particular settings.  In particular, \citet{bolthausen-stein} and \citet{guyon} used Stein's method to prove  central limit theorems for stochastic processes and for random fields, allowing polynomial decay of strong-mixing coefficients.  For random processes where the correlation is due to  shared latent Gaussian random variables, the strong mixing coefficients are bounded by the $\rho$-mixing coefficients, which are in turn bounded by the correlation coefficients in the latent Gaussian variables \citep{bradley-mixing}.

\end{document}